\documentclass[reprint,amsmath,amssymb,aps,prd]{revtex4-2}
\usepackage[colorlinks,citecolor=blue,linkcolor=blue,anchorcolor=blue,filecolor=blue,
urlcolor=blue]{hyperref}
\usepackage{graphicx}% Include figure files
\usepackage{ulem}
\usepackage{dcolumn}% Align table columns on decimal point
\usepackage{bm}% bold math
\usepackage{todonotes}

\begin{document}

\preprint{APS/123-QED}

\title{Structure and astrophysical properties of dark energy admixed neutron stars}
%The central compact object in HESS J1731-347 as a dark energy admixed neutron star

\author{Juan M. Z. Pretel}
 \email{juan04manuel91@gmail.com}
 \affiliation{Centro Brasileiro de Pesquisas F{\'i}sicas, Rua Dr.~Xavier Sigaud, 150 URCA, Rio de Janeiro CEP 22290-180, RJ, Brazil
}

\author{Sergio B. Duarte}
\email{sbd@cbpf.br}
 \affiliation{Centro Brasileiro de Pesquisas F{\'i}sicas, Rua Dr.~Xavier Sigaud, 150 URCA, Rio de Janeiro CEP 22290-180, RJ, Brazil
}

\author{Mariana Dutra}
\email{marianad@ita.br}
 \affiliation{Departamento de F\'isica e Laborat\'orio de Computa\c c\~ao Cient\'ifica Avan\c cada e Modelamento (Lab-CCAM), Instituto Tecnol\'ogico de Aeron\'autica, DCTA, 12228-900, S\~ao Jos\'e dos Campos, SP, Brazil
}

\author{Odilon Lourenço}
\email{odilon.ita@gmail.com}
 \affiliation{Departamento de F\'isica e Laborat\'orio de Computa\c c\~ao Cient\'ifica Avan\c cada e Modelamento (Lab-CCAM), Instituto Tecnol\'ogico de Aeron\'autica, DCTA, 12228-900, S\~ao Jos\'e dos Campos, SP, Brazil
}

\date{\today}% It is always \today, today,
             %  but any date may be explicitly specified

\begin{abstract}
The presence of a dark energy (DE) core, made of a Chaplygin dark fluid, is capable of influencing the diverse observable properties of neutron stars (NSs). However, there is also the possibility that NSs could contain a mixture of ordinary nuclear matter and DE. In this perspective, we propose an equation of state (EoS) that includes both normal matter and DE, where a free parameter $f$ quantifies the ratio of DE density with respect to the total energy density. As expected, in the limit $f \rightarrow 0$ we recover the already known results corresponding to pure nuclear matter. Remarkably, increasing this fraction $f$ leads to a softer EoS and thus strongly favors the fulfillment of the causality condition. As a consequence of our mass--radius results, the central compact object in HESS J1731-347 can be satisfactorily described as a DE admixed NS. For sufficiently high stellar masses, our calculations further indicate that both the gravitational redshift and the fundamental nonradial oscillation frequency increase significantly with an increasing DE fraction, whereas the opposite trend is observed for the total gravitational mass and tidal deformability. As is typical for compact stars composed purely of ordinary matter, we find that in our DE admixed stellar models the central density corresponding to the maximum-mass configuration coincides with the point at which the squared radial vibration frequency vanishes. Moreover, higher DE concentrations lead to enhanced stability of these configurations.
\end{abstract}

%\keywords{Suggested keywords}%Use showkeys class option if keyword
                              %display desired
\maketitle

%\tableofcontents

\section{Introduction}

The Chaplygin gas (CG) and its generalized models provide an important framework into modern cosmology, as they could potentially explain the accelerated expansion of the current universe \cite{Bento2003,Bertolami2004,Barreiro2008,Park2010,Xu2012,Wang2013,Li2019, Yang2019,Mamon2022,Marttens2023}. Initially proposed by the Russian physicist Chaplygin to describe an adiabatic aerodynamic process, the original CG model $p= -B/\rho$ has undergone a series of modifications and generalizations to satisfy observational data coming from various sources including supernovae, the Cosmic Microwave Background (CMB), Baryonic Acoustic Oscillations (BAO) and X-ray measurements of galaxy clusters. The Chaplygin-type equation of state (EoS) introduces negative pressure driving an accelerated stage at late times through DE, and also has a well-established connection with string theory; see~\cite{Ogawa2000} for a discussion of the classical solutions for the CG as $d$-branes. Kamenshchik et al.~\cite{Kamenshchik2001} introduced the Generalized Chaplygin Gas (GCG) model, given by $p= -B/\rho^\alpha$, where $B$ is a positive constant, $0< \alpha \leq 1$, and whose cosmological implications have been investigated in Refs.~\cite{Bento2002, Carturan2003, Biesiada2005}. The GCG model was also extended to the Modified Chaplygin Gas (MCG) \cite{Debnath2004}, characterized by an EoS of the form $p= A\rho -B/\rho^\alpha$, where the free parameters $\{A, B, \alpha\}$ have been measured by means of the \textit{Planck} 2015 CMB anisotropy, type-Ia supernovae (SNe Ia) and observed Hubble parameter data sets \cite{Li2019}. Furthermore, other authors \cite{Zheng2022} have investigated the constraint ability of standard candles (QSO[XUV]+SNe Ia) and standard rulers (QSO[AS]+BAO) on a series of CG models.

The most abundant component of the universe is dark energy~(DE), so it is expected to be present or have an effect on spherically symmetric objects such as compact stars \cite{Lobo2006, Gorini2008PRD, banerjee2020}. In fact, different DE candidates have been shown to alter the global properties and stability of neutron stars~(NSs) \cite{Smerechynskyi2021, ASTASHENOK2023, Pretel2024A, Araujo2024, Pretel2024B, Ventagli2025, SINGH2025}. If we therefore assume that spacetime is filled with CG, it is natural to think about the possible existence of dark-energy stars (DESs) described by a Chaplygin-type EoS \cite{Gorini2008PRD, Bertolami2005, Gorini2009}. Since then, single-phase DESs, made of Chaplygin dark fluid~(CDF), have been intensively studied over the last fifteen years \cite{Rahaman2010, Bhar2018, Errehymy2019, Panotopoulos2020EPJP, Tello2020, Panotopoulos2021, Prasad2021, Pretel2023EPJC, Tudeshki2023, Sunzu2023, Jyothilakshmi2024, Krishna2024}. The study of monophasic DESs was then extended to a hybrid context, where the internal core contains a CDF while the external layer is ordinary matter \cite{Pretel2024A, Pretel2024B}. In particular, it was shown that an increase in the energy density jump, controlled by a parameter $\alpha$, leads to an increase in the radial stability of NSs with a CDF core. Additionally, the dark-energy background in the core of the NS generates theoretical predictions that are potentially consistent with the observational mass--radius measurements of millisecond pulsars from NICER data \cite{Riley2019, Miller2019, Riley2021, Miller2021} and tidal deformability constraints from the GW170817 event \cite{Abbott2018}. 

Very recently, Ref.~\cite{Rutherford2026} showed that current NS mass--radius observations cannot rule out the presence of DE cores described by a CDF, although they can constrain the corresponding parameter space to relatively narrow regions. Indeed, using Bayesian inference and current NICER mass--radius measurements, the authors further showed that the presence of CDF cores can substantially broaden the posterior distributions of NS properties while the observations place strong constraints on some of the corresponding CDF parameters. These results highlight the potential of NS observations as a probe of DE physics in the strong-field regime and emphasize the importance of continuing the study of compact stars containing a DE component. Motivated by these results, we extend our analysis to a mixed scenario. In particular, a crucial open question, to which the present work aims to contribute, is how the EoS of normal nuclear matter is affected by the presence of a CDF component and what consequences this admixture may have for the key astrophysical properties of compact stars.

It is worth mentioning that the CG has also been used to represent a dark fluid surrounding black holes~(BHs). Specifically, the geodesic structure, shadow and optical appearance of a BH immersed in a CDF have been investigated by Li and collaborators \cite{Li2023PRD, Li2024}. Under a geometrothermodynamic formalism for phase transitions, the thermodynamic properties and microstructure of AdS BHs with CDF background were examined in Ref.~\cite{SEKHMANI2024101567}, as well as considering Kaniadakis entropy \cite{SEKHMANI202479}. Meanwhile, other authors have investigated the role of test particles in charged AdS BHs bounded by an exotic fluid comprising a modified Chaplygin EoS \cite{JAVED2025101723}. Furthermore, Ref.~\cite{Mustafa2024} explored the circular orbits and accretion disk around AdS BHs surrounded by a dark fluid with Chaplygin-like EoS.

To achieve our goals, we will use a Chaplygin-type EoS to describe the DE component, while for ordinary nuclear matter we will employ two well-known hadronic models, namely, a relativistic mean-field (RMF) and the Skyrme one. Therefore, this work is organized as follows: in Sec.~\ref{section2} we discuss in detail the different EoSs involved in the study of DE admixed NSs, where a parameter $f$ measures the ratio of the DE density to the total energy density. We will present a detailed analysis of the influence of DE on the microphysics and speed of sound of the stellar mixed fluid. Our numerical results in terms of the mass--radius ($M-R$) relations and $f$ are presented and discussed in Sec.~\ref{section3}. In the same section, we explore the effects of $f$ on gravitational redshift, tidal deformability, and the radial and nonradial oscillation spectrum. Finally, we provide our conclusions and future perspectives in Sec.~\ref{section4}.

%%%%%%%%%%%%%%%%%%%%%%%%%%%%%%%%%%%%%%%%%%%%%%%%%
\section{Equations of state}\label{section2}

\begin{figure*}
\includegraphics[width=8.5cm]{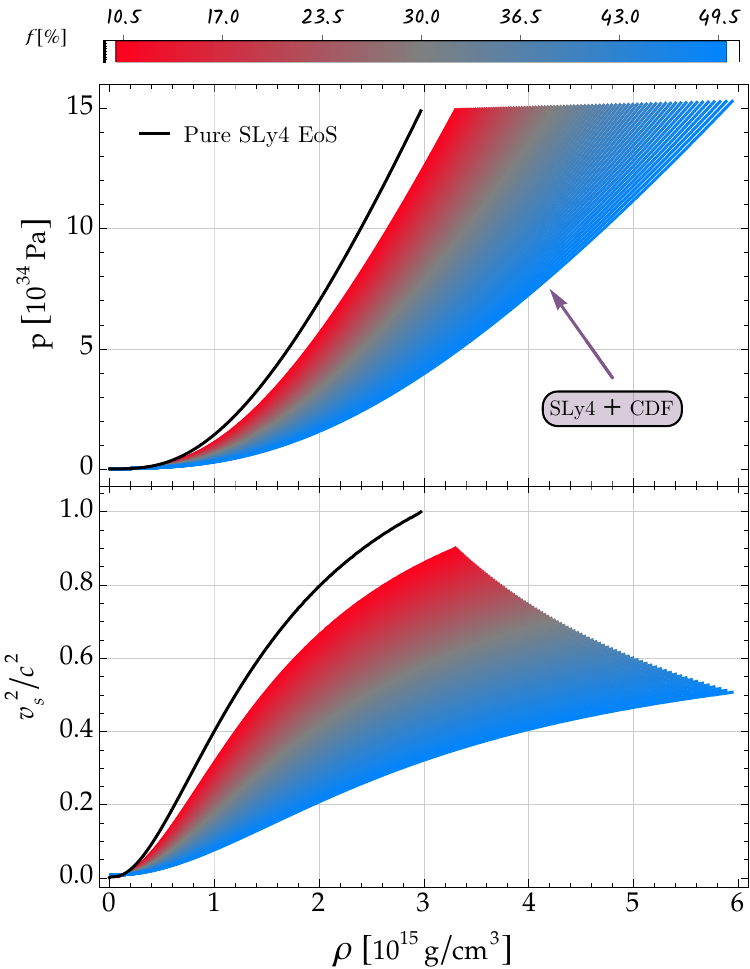}
\hspace{1mm}
\includegraphics[width=8.5cm]{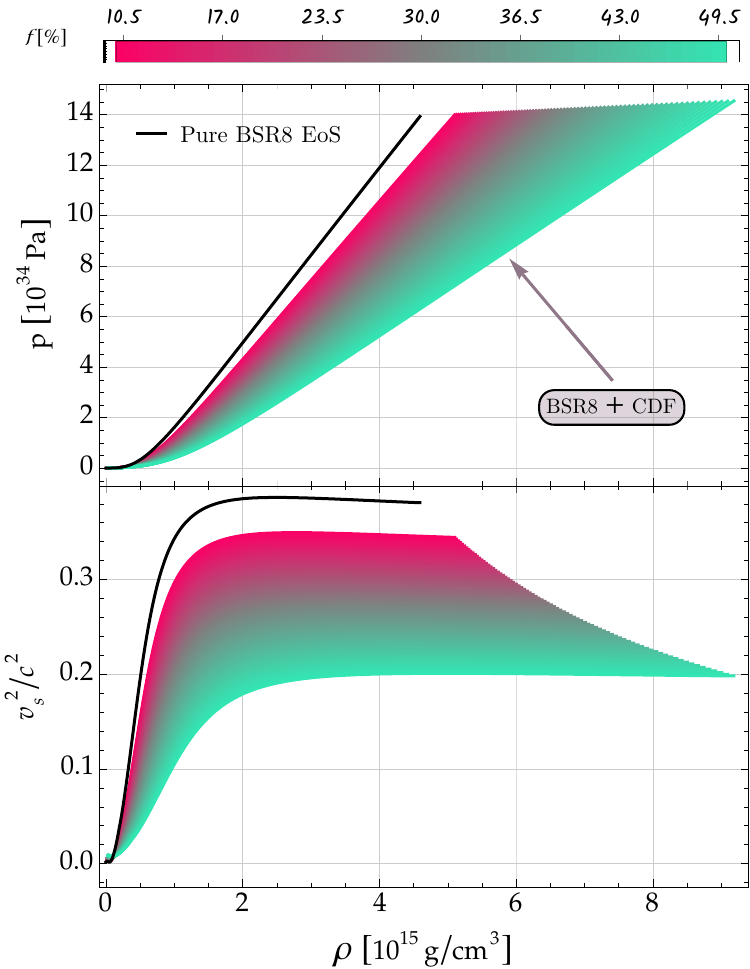}
    \caption{Pressure (upper panels) and square speed of sound (lower panels) as functions of energy density for the mixed stellar fluid considering two hadronic models in the NM sector; the SLy4 EoS (left) and BSR8 EoS (right). In both cases we have varied the percentage fraction $f$ in the range $f \in [10, 50]\%$. Additionally, the black curves provide a representative benchmark for the pure hadronic models, which are recovered when $f \rightarrow 0$. For both models, higher DE concentrations result in a softer total EoS. }%
    \label{FigEoSs}%
\end{figure*}

Our mixed fluid model consists of two components: ordinary nuclear matter and a Chaplygin-type dark fluid. Accordingly, we require two equations of state, each describing the physics of its respective component. A barotropic EoS for dense matter describes the relationship between pressure and density, where pressure is only a function of density. In other words, the microscopic properties of a stellar fluid are determined by its density and pressure, as shown in the form $p= p(\rho)$ or $\rho= \rho(p)$. This EoS is an essential input for the Tolman-Oppenheimer-Volkoff (TOV) equations, thus allowing for the modeling of stellar objects and their behavior in the universe. For normal matter (NM), the EoS is presented in the form of a table that contains a grid of calculated values of density and pressure as $\{\rho_{{\rm NM},i}, p_{{\rm NM},i}\}$. Meanwhile, the Chaplygin-like dark fluid for modeling the DE contribution inside the star is described by
\begin{equation}\label{DEEoS}
    p_{\rm DE} = A\rho_{\rm DE} - \frac{B}{\rho_{\rm DE}}\, ,
\end{equation}
where $\rho_{\rm DE}$ and $p_{\rm DE}$ are the energy density and pressure of the DE fluid, respectively. In this work, we adopt parameters $A$ and $B$ such that they allow us to obtain appreciable changes in our radii and mass results, but also always lead to satisfy causality within the entire mixed stellar fluid. With this in mind, in our numerical calculations we use $A= 0.015$ and $B= 1.0\times 10^{-26}\, \rm m^{-4}$; values given within a geometric unit system. However, we must point out that our final results are given in physical units for the sake of comparison.

To describe the EoS for the normal matter component, we adopt two widely used effective hadronic models: RMF theory~\cite{baoanli08,lattimer24,dutra14} and the nonrelativistic Skyrme-type interaction~\cite{lattimer24,stone07,dutra12}. Both frameworks enable a microscopic derivation of the pressure and energy density on the basis of fundamental nucleon-nucleon interactions by means of the mean-field approach. Their parameters are calibrated to reproduce properties of nuclear matter at saturation and observables related to finite nuclei, as well as neutron star observational data.

The RMF model describes the nucleon as a Dirac field interacting through a meson exchange mechanism, in which the scalar meson $\sigma$ provides attraction, the vector meson $\omega$ is responsible for repulsion, and the isovector meson $\rho$ governs isospin asymmetry. Including nonlinear self-interactions and mixed meson coupling terms improves the model's agreement with empirical data, especially regarding nuclear incompressibility and symmetry energy. From the corresponding energy-momentum tensor, in the mean-field approximation, energy density and pressure can be written as~\cite{baoanli08,lattimer24,dutra14}
\begin{align}
\rho_{\text{RMF}} &= \frac{1}{2}m_\sigma^2\sigma^2 + \frac{\mathcal{A}}{3}\sigma^3 + \frac{\mathcal{B}}{4}\sigma^4 - \frac{1}{2}m_\omega^2\omega_0^2 - \frac{C}{4}(g_\omega^2\omega_0^2)^2 
\nonumber\\
&- \frac{1}{2}m_\rho^2 b_{0(3)}^2 + g_\omega \omega_0 n + \frac{g_\rho}{2} b_{0(3)} n_3
\nonumber\\
&- g_\sigma g_\omega^2 \sigma \omega_0^2 \left( \alpha_1 + \frac{1}{2} \alpha_1' g_\sigma \sigma \right) 
\nonumber \\
&- g_\sigma g_\rho^2 \sigma b_{0(3)}^2 \left( \alpha_2 + \frac{1}{2} \alpha_2' g_\sigma \sigma \right)
- \frac{1}{2} \alpha_3' g_\omega^2 g_\rho^2 \omega_0^2 b_{0(3)}^2 \nonumber \\
&+ \frac{1}{\pi^2}\sum_{i=p,n} \int_0^{k_{F_{i}}}\hspace{-0.2cm}dk\, k^2 \sqrt{k^2 + {M^*}^2},
\end{align}
and
\begin{align}
p_{\text{RMF}} &= - \frac{1}{2}m_\sigma^2\sigma^2 - \frac{\mathcal{A}}{3}\sigma^3 - \frac{\mathcal{B}}{4}\sigma^4 + \frac{1}{2}m_\omega^2\omega_0^2 + \frac{C}{4}(g_\omega^2\omega_0^2)^2 
\nonumber \\
&+ \frac{1}{2}m_\rho^2 b_{0(3)}^2 + g_\sigma g_\rho^2 \sigma b_{0(3)}^2 \left( \alpha_2 + \frac{1}{2} \alpha_2' g_\sigma \sigma \right)
\nonumber \\
&+ \frac{1}{2} \alpha_3' g_\omega^2 g_\rho^2 \omega_0^2 b_{0(3)}^2 + g_\sigma g_\omega^2 \sigma \omega_0^2 \left( \alpha_1 + \frac{1}{2} \alpha_1' g_\sigma \sigma \right) 
\nonumber \\
&+ \frac{1}{3\pi^2}\sum_{j=p,n}\int_0^{k_{F_{j}}}\hspace{-0.3cm}\frac{dk\,k^4}{\sqrt{k^2 + M^{*2}}},
\end{align}
respectively, with $n=n_p+n_n$ and $n_3=n_p-n_n$. The effective nucleon mass is $M^* = M_{\text{nuc}} - g_\sigma \sigma$ and $k_{F_j}$ is the Fermi momentum of the nucleon $j$, related to the corresponding density through ${k_F}_{j}=(3\pi^2n_{i})^{1/3}$. The mean-field mesons fields are $\sigma$, $\omega_0$ and $b_{0(3)}$, and $M_{\text{nuc}}$ is the nucleon rest mass.

In contrast, the Skyrme model provides a nonrelativistic approach based on a zero-range effective interaction with density-dependent and momentum-dependent terms that simulate many-body effects. The energy density and pressure read~\cite{lattimer24,stone07,dutra12}
\begin{align}
\rho_{\text{Sky}} &= \frac{3}{10M_{\text{nuc}}}\left(\frac{3\pi^2}{2}\right)^{2/3}n^{5/3}H_{5/3} 
+ \frac{t_0}{8}n^2[2(x_0+2)
\nonumber \\
&-(2x_0+1)H_2] 
+ \frac{t_3}{48}n^{\sigma+2} [2(x_3+2)-(2x_3+1)H_2] \nonumber\\
&+ \frac{3}{40}\left(\frac{3\pi^2}{2}\right)^{2/3}n^{8/3}\left(aH_{5/3}+bH_{8/3}\right) + nM_{\text{nuc}},
\end{align}
and
\begin{align}
p_{\text{Sky}} &= \frac{1}{5M_{\text{nuc}}}\left(\frac{3\pi^2}{2}\right)^{2/3}n^{5/3}H_{5/3} 
\nonumber\\
&+ \frac{t_0}{8}n^2[2(x_0+2) -(2x_0+1)H_2] 
\nonumber\\
&+ \frac{t_3}{48}(\sigma+1)n^{\sigma+2}
[2(x_3+2)-(2x_3+1)H_2] \nonumber\\
&+ \frac{1}{8}\left(\frac{3\pi^2}{2}\right)^{2/3}n^{8/3}\left(aH_{5/3}+bH_{8/3}\right),
\end{align}
where
\begin{align}
a &= t_1(x_1+2)+t_2(x_2+2), \\
b &= \frac{1}{2}[t_2(2x_2+1)-t_1(2x_1+1)], \\
H_\ell(y) &= 2^{\ell-1}[y^\ell+(1-y)^\ell],
\end{align}
and $y = n_p/n$ is the proton fraction. For stellar matter, leptons are added to ensure charge neutrality, and beta-equilibrium is imposed. See equations (14) and (15) of Ref.~\cite{Pretel2024B}. The Baym, Pethick, and Sutherland (BPS) EoS~\cite{bps} is added to describe the crust at low densities. Both the RMF and Skyrme models provide EoSs in tabulated form \(\{ \rho_{{\rm NM},i}, p_{{\rm NM},i} \}\), where each pair corresponds to a thermodynamic state of the hadronic component. In our framework, the index NM refers either to the RMF-based description with the BSR8~\cite{bsr8} set or to the Skyrme model with SLy4~\cite{sly4} parametrization, including leptons and BPS. These parametrizations were selected for their ability to accurately reproduce various nuclear properties, including charge radii, binding energies of ground states, and giant monopole resonances for a set of spherical nuclei: $^{16}\rm O$, $^{34}\rm Si$, $^{40}\rm Ca$, $^{48}\rm Ca$, $^{52}\rm Ca$, $^{54}\rm Ca$, $^{48}\rm Ni$, $^{56}\rm Ni$, $^{78}\rm Ni$, $^{90}\rm Zr$, $^{100}\rm Sn$, $^{132}\rm Sn$, and $^{208}\rm Pb$. Furthermore, they are also consistent with some global characteristics of neutron stars. A comprehensive analysis involving 415 parametrizations of both, the RMF model and the Skyrme one can be found in Ref.~\cite{brett-jerome}.
%%%%%%%%%%%%%%%%%%%%%%%%%%%%%%%%%%%%%%%%  

The procedure for determining the total EoS, $p = p(\rho)$, describing the mixed fluid is as follows: given an energy density value $\rho_{{\rm NM},i}$ from the tabulated data of the NM fluid, we determine the dark energy density using the relation 
\begin{eqnarray}
\rho_{{\rm DE},i} = f(\rho_{{\rm DE},i} + \rho_{{\rm NM},i}).
\end{eqnarray}
This means that the new parameter $f$ allows us to set the ratio of the DE density to the total energy density of the mixed fluid. It is worth emphasizing that, in the vanishing-$f$ case, the DE density does not contribute and the total fluid becomes purely hadronic. Consequently, the Chaplygin-like EoS \eqref{DEEoS} ceases to be physically meaningful. Indeed, such equation becomes ill-defined when $f=0$, since the second term on the right-hand side diverges. Nevertheless, for sufficiently small but nonzero values of $f$, the purely hadronic configurations can still be approximately recovered. Finally, after finding $\rho_{{\rm DE},i}$, we proceed to calculate the DE pressure according to Eq.~\eqref{DEEoS}. As a result, our total EoS $\{\rho_i, p_i\}$ is constructed from the following grid of calculated values
\begin{equation}\label{TotalEoSEq}
    {\rm Mixed\ fluid\ EoS} : 
  \begin{cases}
    \rho_i = \rho_{{\rm DE},i} + \rho_{{\rm NM},i}\,  \\ 
    p_i = p_{{\rm DE},i} + p_{{\rm NM},i}\,
  \end{cases}
\end{equation}
where the total energy density and pressure are the sums of its individual fluid components. These tabulated points are then interpolated so that we obtain a one-parameter EoS of the form $p= p(\rho)$ and thus solve the stellar structure equations.

The EoSs for our mixed stellar fluid are shown in the upper plots of Fig.~\ref{FigEoSs}, where we have adopted a wide range of values for $f$. The left and right panels correspond to the SLy4 and BSR8 EoSs describing the hadronic material content, respectively. As discussed above, when the DE fraction becomes negligible compared to the total stellar content, the EoS reduces to the purely hadronic case, which is represented by the black lines in both scenarios. The increase in $f$ leads to a softer total EoS compared to the pure nuclear model, and as we will see later, this has a substantial effect on the $M-R$ diagrams.

The speed of sound, which represents the speed at which disturbances propagate through a stellar medium, is a crucial factor in determining causality. Specifically, the causality condition states that the speed of sound $v_s$ must not exceed the speed of light $c$, that is, $v_s<c$. The lower panels of Fig.~\ref{FigEoSs} indicate that the square speed of sound, defined by $v_s^2= dp/d\rho$, decreases as dark energy is supplemented into the dense stellar fluid. In other words, the main consequence of increasing $f$ is to significantly decrease the speed of sound. One observes that both NM models do not allow for superluminal signal propagation, so that the causality condition is always respected throughout the whole range of energy densities adopted. This result is of great importance because some EoSs provided in the literature do not support causality but could become causal in the presence of DE. Specifically, the speed of sound predicted by some versions of the APR EoS \cite{APRref} can exceed the speed of light, particularly in some high-density regimes, but could be causal if DE effects are taken into account. This is also the case for the nonrelativistic Skyrme model used here. Notice that for $f=0$ (no DE content), SLy4 reaches $v_s/c \simeq 1$ at $\rho \simeq 2.97\times 10^{15}\, \mathrm{g/cm^3}$. However, for $f>0$, the energy-density interval over which $v_s/c < 1$ becomes larger.

%%%%%%%%%%%%%%%%%%%%%%%%%%%%%%%%%%%%%%%%%%%%%%%%%
\section{Global properties}\label{section3}

\begin{figure*}
\includegraphics[width=8.5cm]{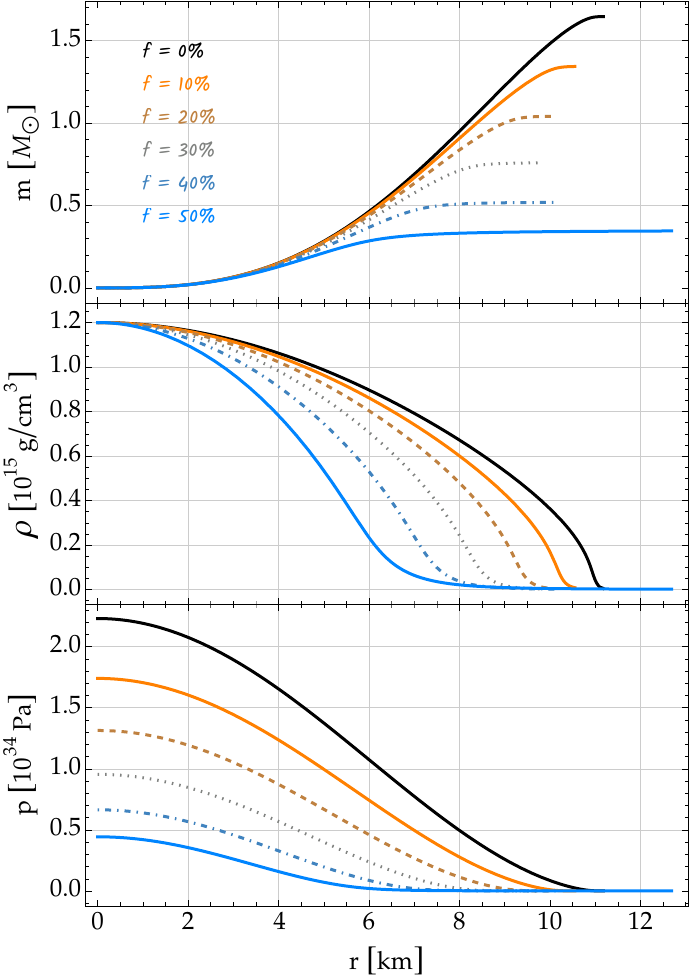}
\hspace{1mm}
\includegraphics[width=8.5cm]{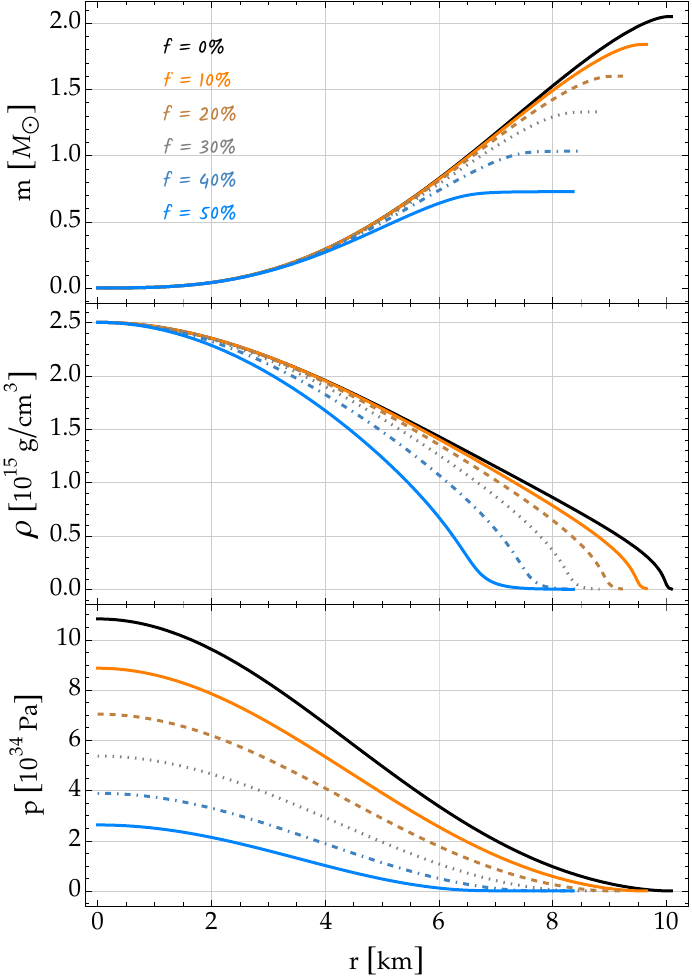}
    \caption{Radial behavior of the mass function, energy density and pressure assuming a ${\rm SLy4+CDF}$ mixed fluid with $f\in [0,50]\%$ for two central density values; $\rho_c= 1.2\times 10^{15}\, \rm g/cm^3$ (left) and $\rho_c= 2.5\times 10^{15}\, \rm g/cm^3$ (right). An increase in the DE concentration within the stellar interior consistently leads to a decrease in the mass function, whereas the surface radius $R$ can either decrease or increase depending on the central energy density. Note also that, for variations of $f$, both the energy density and the pressure decrease monotonically with the radial coordinate. }%
    \label{FigRadBehav}%
\end{figure*}

\subsection{Mass--radius diagrams and gravitational redshift}

The most fundamental global properties of a NS are its mass and radius~\cite{Lattimer2004}, which are determined by solving the relativistic equations of hydrostatic equilibrium, namely the Tolman–Oppenheimer–Volkoff (TOV) equations~\cite{Tolman:1939jz, Oppenheimer:1939ne}:
\begin{align}
    \frac{dm}{dr} &= 4\pi r^2\rho ,  \label{TOV1}  \\
    \frac{dp}{dr} &= -(\rho + p)\left[ \frac{m}{r^2} + 4\pi r p \right]\left( 1 - \frac{2m}{r} \right)^{-1} ,  \label{TOV2}
\end{align}
where the function $m(r)$ represents the gravitational mass enclosed within the radial coordinate $r$. These equations result from the Einstein field equations applied to a static, spherically symmetric metric with a matter-energy distribution described by an isotropic perfect fluid. The EoS, $p = p(\rho)$, closes the system and allows the determination of $\rho(r)$ and $m(r)$ using only two initial conditions: $\rho(0) = \rho_c$ and $m(0) = 0$, where $\rho_c$ denotes the central energy density. Once the tabulated points in the mixed fluid EoS \eqref{TotalEoSEq} are interpolated, we proceed to numerically solve the stellar structure differential equations \eqref{TOV1} and \eqref{TOV2} for a given central density $\rho_c$. For the ${\rm SLy4+CDF}$ mixed fluid, Fig.~\ref{FigRadBehav} shows the numerical solutions for two values of $\rho_c$, where we have varied $f$ between $0$ and $50\%$. The left panel corresponds to $\rho_c= 1.2\times 10^{15}\, \rm g/cm^3$, while the right panel is obtained for a higher central density $\rho_c= 2.5\times 10^{15}\, \rm g/cm^3$. Regardless of the value of $f$, the energy density and pressure are decreasing functions, while the mass $m(r)$ increases as one moves away from the stellar center.

The surface radius of the star $R$ is calculated when the pressure vanishes, i.e., $p(r=R) =0$. Once the radius is known, the total gravitational mass of the compact star is simply given by $M= m(R)$. According to Fig.~\ref{FigRadBehav}, we can observe that $R$ always decreases with increasing $f$ for high central densities (top right panel), while for a low central density (top left panel), the radius behavior is less trivial; it can decrease and then increase as the DE concentration varies. From the left panel, we can state that for low central densities and high $f$ (see, for example, the case with $f= 50\%$) most of the energy and mass are contained within the sphere with radius $\sim R/2$ while there is a large envelope whose mass-energy contributions are negligible. Since the central energy density plays a crucial role in determining the radius of our stellar configurations under the effects of $f$, we will later analyze in more detail the dependence of $R$ on the central density for different values of the parameter $f$.

\begin{figure*}
\includegraphics[width=8.4cm]{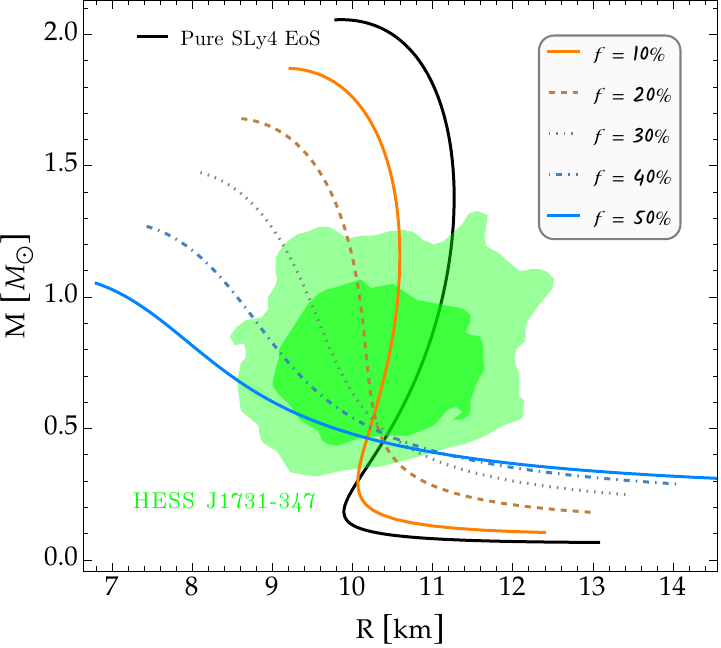}
\hspace{1mm}
\includegraphics[width=8.4cm]{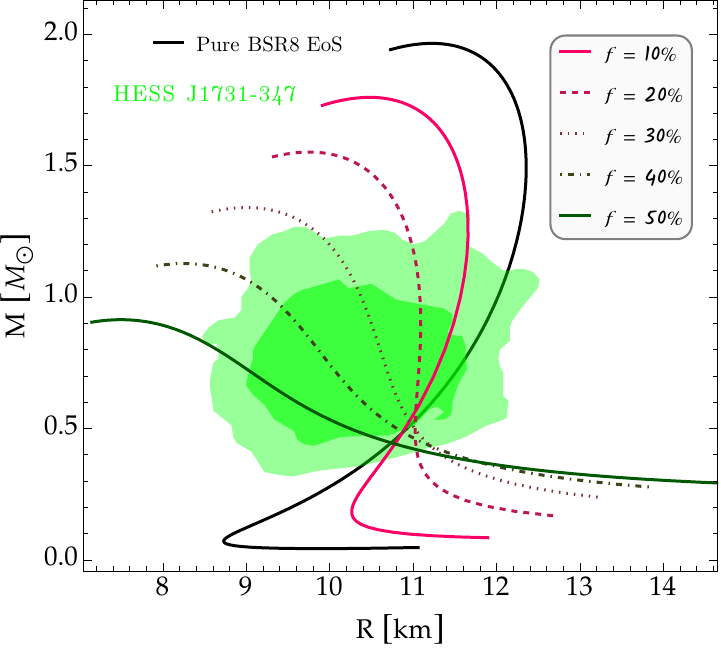}
    \caption{Mass--radius diagram for dark energy admixed neutron stars for several values of $f$, where we have also included the pure nuclear case by black curves. The green filled areas represent the $1\sigma$ and $2\sigma$ confidence levels for the supernova remnant HESS J1731-347 \cite{Doroshenko2022}. }%
    \label{FigMR}%
\end{figure*}

Each value of central energy density produces a DES with a certain radius $R$ and mass $M$. Consequently, the variation of the central energy density given on the horizontal axes of Fig.~\ref{FigEoSs} leads us to obtain the so-called $M-R$ diagrams. For the two hadronic models, our $M-R$ results are presented in Fig.~\ref{FigMR}, where we have considered five values for the percentage fraction $f$. For comparison purposes, the pure nuclear scenario has been included by black curves, which actually correspond to $f=0$, i.e., when there is no presence of DE. In the high-mass branch, we observe that the mass decreases drastically with increasing $f$, while the effect reverses at small masses. In other words, high concentrations of DE within a NS cause a decrease in maximum mass values up to $\sim 1\, M_\odot$ when $f= 50\%$. Of course, this favors the description of low-mass compact stars, such as the central compact object within the supernova remnant HESS J1731-347 \cite{Doroshenko2022}. It has been conjectured that this object is either the lightest NS known, or a ``strange star'' with a more exotic EoS. In that regard, here we point out that such a supernova remnant may be highly consistent with our hypothesis that a NS could contain hadronic material mixed with DE.

\begin{figure*}
\includegraphics[width=8.4cm]{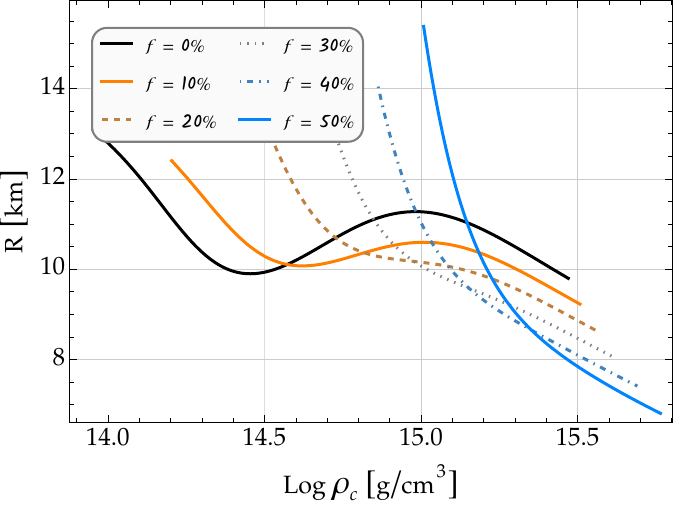}
\hspace{1mm}
\includegraphics[width=8.4cm]{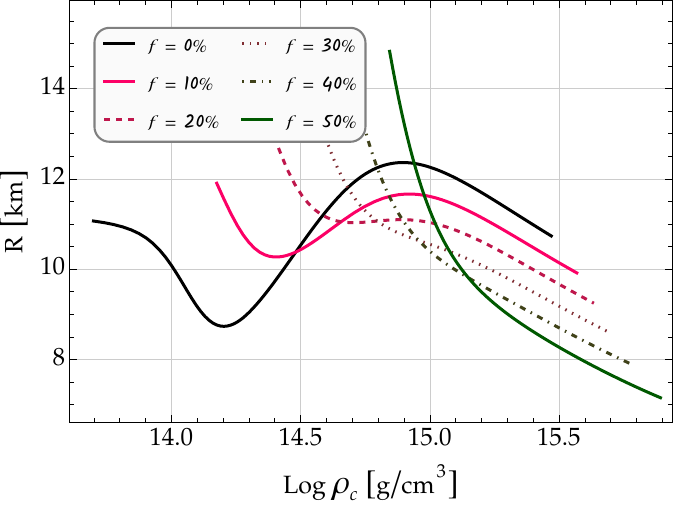}
    \caption{Radius versus central density for our DE admixed NS models under the effect of the percentage parameter $f$, where ordinary matter is described by the SLy4 EoS (left) and BSR8 EoS (right). }%
    \label{FigRadiusDen}%
\end{figure*}

As previously mentioned, we now investigate the dependence of the stellar radius on the central density under the influence of the fraction $f$ (see Fig.~\ref{FigRadiusDen}). For low DE concentrations (e.g., $f = 10\%$), the radius exhibits a behavior similar to that of the pure SLy4 scenario (black curve): $R$ decreases, then increases, and finally decreases again as $\rho_c$ increases.  In contrast, for larger values of $f$, the radius decreases monotonically with increasing $\rho_c$.

Another important astrophysical observable is the surface gravitational redshift $Z_g$, a relativistic effect by which radiation emitted from the stellar surface is shifted toward longer wavelengths as it escapes the star’s gravitational field. Such an observable provides a test of General Relativity in the strong-field regime and serves as a key tool for constraining the stellar mass, radius, and the dense-matter EoS within the stellar interior \cite{Lindblom1984, Hambaryan2017, Tang2020, Luo2022}. Furthermore, recent studies have shown that $Z_g$ can satisfy universal relations \cite{Yang2022, Chatterjee2025}, namely, correlations that are approximately independent of the EoS and relate it to other macroscopic quantities, such as mass, radius, or tidal deformability. In other words, these universal relations reduce the sensitivity to the microphysics of dense matter and render the gravitational redshift a particularly valuable observable for inferring the global properties of compact stars from astrophysical observations. Motivated by these studies, we calculate $Z_g$ for our DESs while incorporating observational measurements. This relativistic observable is defined as
\begin{equation}
    Z_g = \frac{1}{\sqrt{1- 2M/R}} -1 ,
\end{equation}
and is displayed in the upper plots of Fig.~\ref{FigzLfvsMass} for both NM models within the mixed fluid. At high $Z_g$, our results indicate that the primary effect of increasing $f$ is an enhancement of the gravitational redshift for a given stellar mass $M$. Redshift measurements for the RX J1856.5-3754, RX J0720.4-3125 and RBS 1223 sources have been included in our plots for comparison purposes, whose redshifts are $0.22_{-0.12}^{+0.06}$, $0.205_{-0.003}^{+0.006}$ and $0.16_{-0.02}^{+0.03}$, respectively \cite{Hambaryan2017, Tang2020, Chatterjee2025}. Figure \ref{FigzLfvsMass} confirms the consistency between the observational measurements and our theoretical redshift predictions.

\subsection{Tidal deformability and nonradial oscillations}

In order to compare our findings with gravitational-wave (GW) observations, we will also calculate the dimensionless tidal deformability $\Lambda= 2k_2/3C^5$, where $k_2$ is the Love number and $C= M/R$ is the compactness. In fact, tidal deformability is a key observable in the GW signal emitted during the inspiral phase of binary NS mergers \cite{Most2018, Raithel2019, Chatziioannou2020, Dietrich2021}, as it imprints characteristic signatures that allow direct constraints on the EoS and the internal structure of these compact stars. The Love number can be obtained after finding the tidal perturbation $y(r)$ by the differential equation \cite{Hinderer2010}
\begin{equation}\label{yEq}
    r\frac{dy}{dr} +y^2 + \mathcal{F}y + r^2\mathcal{H} =0 ,
\end{equation}
where
\begin{align}
    \mathcal{F} =&\ e^{2\Psi}\left[ 1 + 4\pi r^2(p - \rho) \right] ,  \\
    \mathcal{H} =&\ 4\pi e^{2\Psi}\left[ 5\rho + 9p + \frac{\rho+ p}{v_s^2} \right] - \frac{6e^{2\Psi}}{r^2} - 4\Phi'^2 .
\end{align}

\begin{figure*}
\includegraphics[width=8.6cm]{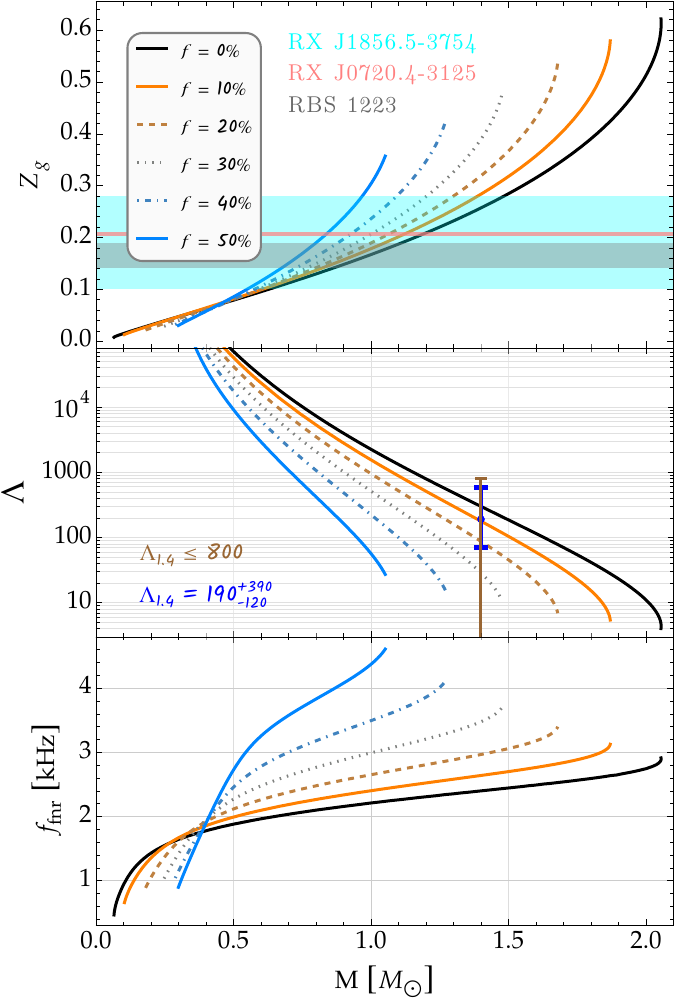}
\hspace{1mm}
\includegraphics[width=8.6cm]{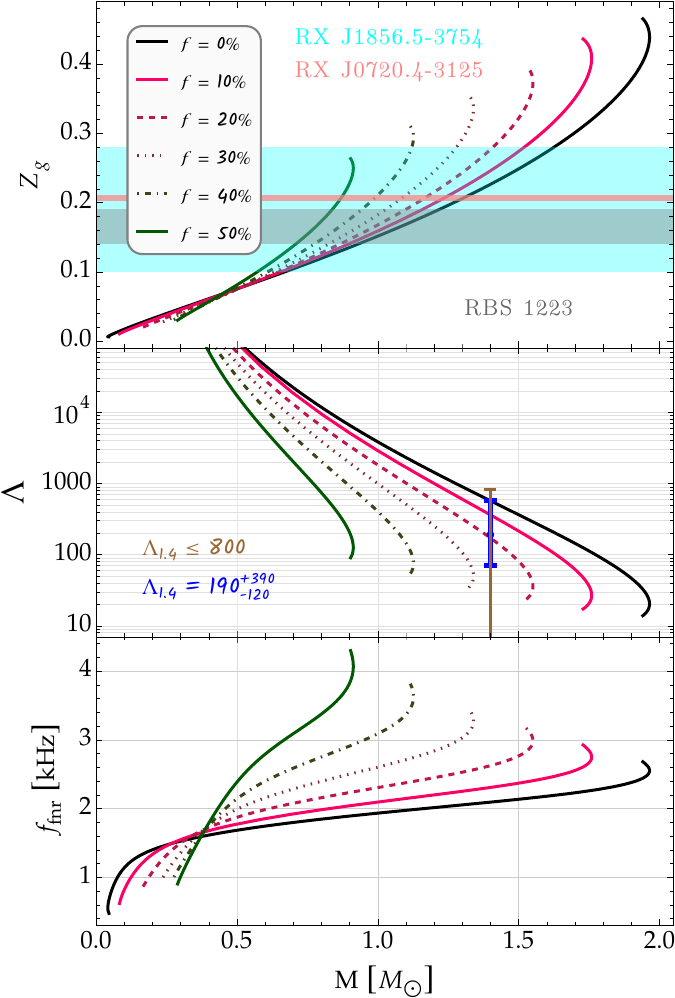}
    \caption{Gravitational redshift (top), tidal deformability (middle), and fundamental nonradial mode frequency (bottom) as a function of mass for the stellar configurations presented in Fig.~\ref{FigMR}. The top panels include redshift observational measurements for the RX J1856.5-3754, RX J0720.4-3125, and RBS 1223 sources \cite{Hambaryan2017, Tang2020, Chatterjee2025}. The blue vertical line in the middle panels denotes $\Lambda_{1.4}= 190_{-120}^{+390}$ reported by LIGO-Virgo Collaboration from the GW170817 signal \cite{Abbott2018}, while the brown line represents the EoS-independent bound $\Lambda_{1.4} \leq 800$ from Ref.~\cite{Abbott2017PRL}. It is observed that, for stellar masses above approximately $0.4\, M_\odot$, both the gravitational redshift and the $f$-mode frequency increase with increasing DE presence, while the tidal deformability decreases for a fixed mass $M$. The hadronic part of each star is described by SLy4 (left panels) and BSR8 (right panels) parametrizations.}
    \label{FigzLfvsMass}%
\end{figure*}

Note that the metric variables ($\Psi$ and $\Phi$) and thermodynamic quantities ($\rho$ and $p$) are obtained by solving the TOV equations. Using these solutions, together with the boundary condition $y(0)=2$ \cite{Hinderer2010, Postnikov2010}, we obtain $y(r)$ from Eq.~\eqref{yEq}. This allows us to calculate the Love number of our DE admixed NSs via
\begin{align}\label{LoveNumEq}
    k_2 =&\ \frac{8}{5}(1- 2C)^2C^5 \left[ 2C(\alpha -1) - \alpha+ 2 \right]  \nonumber  \\
    &\times \left\lbrace 2C[ 4(\alpha+ 1)C^4 + (6\alpha- 4)C^3 \right.  \nonumber  \\
    &\left.+\ (26- 22\alpha)C^2 + 3(5\alpha -8)C - 3\alpha+ 6 \right]   \nonumber  \\
    &\left.+\ 3(1-2C)^2\left[ 2C(\alpha- 1)- \alpha +2 \right]\ln(1-2C) \right\rbrace^{-1} ,
\end{align}
with $\alpha= y(R)- 4\pi R^3 \rho_s/M$ and $\rho_s$ represents the energy density at the surface of the mixed fluid model \cite{Yang2023}. For the stellar configurations presented in Fig.~\ref{FigMR}, the dimensionless tidal deformability as a function of the gravitational mass is shown in the middle panels of Fig.~\ref{FigzLfvsMass}, where we included the LIGO–Virgo Collaboration estimate of the tidal effects from the GW170817 signal \cite{Abbott2018}, namely $\Lambda_{1.4}= 190_{-120}^{+390}$ (vertical blue line). However, we emphasize that this constraint relies on the assumption that the coalescing objects were standard NSs. In the case of the DE admixed NSs considered in the present study, it is more appropriate to adopt the EoS–independent bound $\Lambda_{1.4} \leq 800$ (vertical brown line) reported in the earlier LIGO–Virgo analysis \cite{Abbott2017PRL}, which is satisfied for $f\lesssim 30\%$. Our results further demonstrate the significant impact of DE on the tidal deformability of DESs. For a fixed stellar mass $M$, the tidal deformability $\Lambda$ decreases markedly as the DE fraction $f$ increases.

In the Cowling approximation \cite{Sotani2011}, the calculation of nonradial oscillation frequencies, in particular the fundamental mode frequency, is of special importance because this mode is strongly correlated with the global properties of a compact star \cite{Andersson1998, Lau2010, chan2014multipolar, Pretel2024} while being relatively insensitive to the gravitational perturbations neglected in this approximation. The $f$-mode also provides the dominant contribution to GW emission in dynamical processes, making it a potentially observable quantity. Consequently, even within the simplified Cowling framework, the study of the fundamental mode yields valuable insight into the internal structure and EoS of ultradense matter, serving as an effective link between theoretical modeling and astrophysical observations. This motivates us to also study nonradial oscillations in our stellar models and analyze the implications of DE on the fundamental nonradial oscillation frequency ($f_{\rm fnr}$), which is the lowest-order nonradial mode and exhibits no nodes. The nonradial perturbations of relativistic compact stars in the Cowling approximation are described by the following differential equations \cite{Sotani2011, Kumar2024, Zhang2024PRD}
\begin{align}
    \frac{dW}{dr} &= \frac{1}{v_s^2}\left[ \frac{\omega_{\rm nr}^2r^2 V}{e^{2\Phi- \Psi}} + \Phi'W\right] - \ell(\ell+1)e^\Psi V,  \label{NonRadEq1}  \\
    \frac{dV}{dr} &= 2\Phi'V - \frac{e^\Psi W}{r^2},  \label{NonRadEq2}
\end{align}
where $W(r)$ and $V(r)$ are perturbative variables, and $\omega_{\rm nr}$ is the nonradial oscillation frequency to be determined. The system of equations above is integrated from the stellar center to the surface. At the center, regularity conditions impose
$W = a r^{\ell+1}$ and $V = -a r^{\ell}/\ell$, where the proportionality constant $a$ simply fixes the overall normalization of the nonradial pulsation modes, while $\ell$ is the spherical harmonic index. For nonradial oscillations, one always has $\ell\geq 1$, however, $\ell=2$ is the most astrophysically relevant case, since quadrupole modes couple directly to GW emission. At the surface, the requirement that the Lagrangian pressure perturbation vanishes leads to the following boundary condition:
\begin{equation}\label{SurCond_NRO}
    \frac{\omega_{\rm nr}^2 V}{e^{2\Phi}} + \frac{\Phi'W}{e^{\Psi}r^2} =0 .
\end{equation}

We solve this eigenvalue problem using the well-known shooting method. Specifically, Eqs.~\eqref{NonRadEq1} and \eqref{NonRadEq2} are integrated for a set of trial values of the nonradial pulsation frequency $\omega_{\rm nr}$, and the physically admissible eigenfrequencies are identified as those satisfying the boundary condition \eqref{SurCond_NRO} at $r=R$. In the lower panels of Fig.~\ref{FigzLfvsMass}, we display the fundamental nonradial oscillation frequency $f_{\rm fnr} = \omega_{\rm fnr}/(2\pi)$ as a function of the gravitational mass. For a fixed DE fraction $f$, the frequency $f_{\rm fnr}$ increases monotonically with the stellar mass $M$. On the other hand, for masses $\gtrsim 0.4\,M_\odot$, $f_{\rm fnr}$ increases substantially with increasing DE concentration at fixed $M$. Our findings show that DE leaves a distinct and potentially observable imprint on the nonradial oscillation spectra of NSs. There is currently no direct observational detection of the fundamental nonradial mode frequency from NSs. Nevertheless, analyzes of GW data, most notably those based on the GW170817 event \cite{Wen2019, Pratten2020}, have enabled indirect constraints on the plausible range of $f$-mode frequencies by exploiting asteroseismology and quasi-universal relations that link the mode frequency to macroscopic stellar properties. These studies indicate that, for ordinary hadronic NSs, the $f$-mode frequency is expected to lie in the kHz range, typically around $1.5\text{--}3\,\rm{kHz}$. Remarkably, future generations of GW detectors are expected to reach the sensitivity required to directly measure $f$-mode frequencies, thus providing a powerful observational probe of the NS structure and of possible exotic components, such as DE, in their interiors. Consequently, such imprints offer a strong discriminant between purely hadronic stars and DESs, making the $f$-mode an exceptionally effective probe of DE effects.

\begin{figure*}
\includegraphics[width=8.4cm]{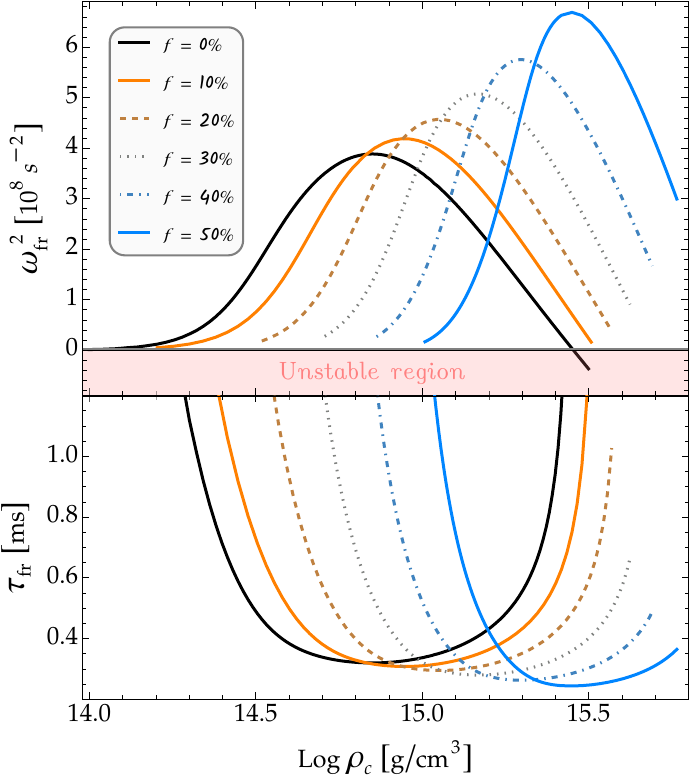}
\hspace{1mm}
\includegraphics[width=8.4cm]{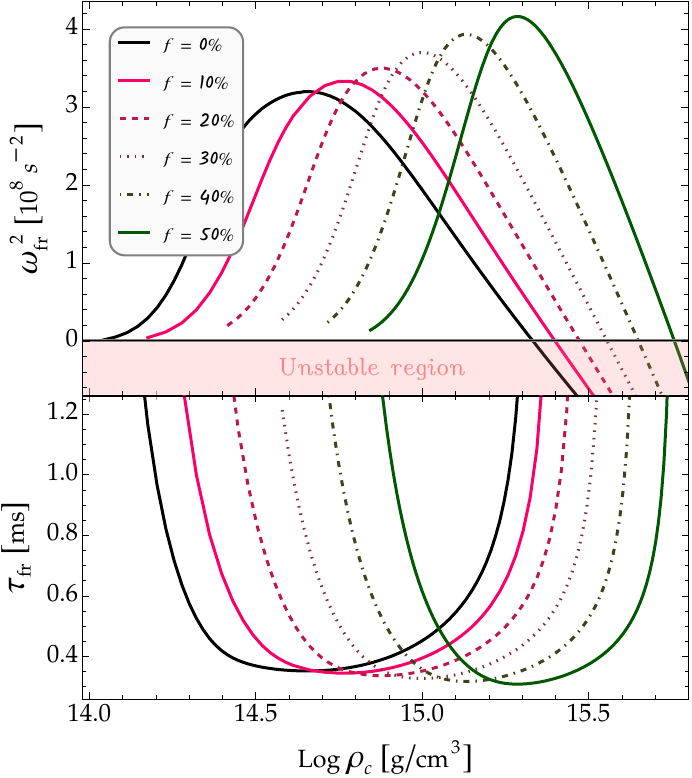}
    \caption{Radial oscillation spectrum for our stellar DE admixed configurations, where the NM component is described by the SLy4 EoS (left) and BSR8 EoS (right). The upper panels show the squared frequency of the fundamental radial pulsation mode $(\omega^2_{\rm fr})$, while the lower panels show the corresponding radial oscillation period $(\tau_{\rm fr})$, both as functions of the central density $\rho_c$. The pink regions represent imaginary oscillation frequencies, where compact stars are dynamically unstable under small radial perturbations. }%
    \label{FigRadFreq}%
\end{figure*}

\subsection{Radial oscillations and stellar stability}

Nonradial oscillations in compact stars break spherical symmetry and induce fluid angular deformations, which being particularly relevant because modes with $\ell\geq2$ generate GW emission and encode direct information about the global properties of the star. By contrast, radial oscillations, although not observable through GWs, play a crucial role in the analysis of dynamical stability of a compact star against collapse or expansion, since the fundamental radial mode generally vanishes at the maximum-mass configuration. For this reason, the study of the radial oscillation spectrum is essential to identify the stability regime of stellar configurations and to ensure that the nonradial models considered correspond to physically stable equilibrium stars \cite{ChandrasekharApJ, ChandrasekharPRL}. 

In particular, the squared frequency of the fundamental radial mode $\omega_{\rm fr}^2$ serves as a direct stability indicator; configurations with $\omega_{\rm fr}^2>0$ are dynamically stable, $\omega_{\rm fr}^2=0$ marks the onset of marginal stability (typically at the maximum-mass configuration) and $\omega_{\rm fr}^2<0$ signals instability toward gravitational collapse. With $\zeta$ defined in terms of the Lagrangian displacement $\xi$ as $\zeta = \xi/r$, and $\Delta p$ denoting the Lagrangian perturbation of the fluid pressure, the adiabatic radial pulsations of compact stars in Einstein gravity are governed by the following system of first-order differential equations \cite{Zhang2024PRD, Gondek1997, Vasquez2010, Pretel2020MNRAS, Bora2021, Hong2023}:
\begin{align}
    \frac{d\zeta}{dr} &= \mathcal{J}\zeta + \mathcal{K}\Delta p ,    \label{ROEq1}  \\
    \frac{d(\Delta p)}{dr} &= \mathcal{X}\zeta + \mathcal{Y}\Delta p ,  \label{ROEq2}
\end{align}
with the coefficients given by
\begin{align}
    \mathcal{J} &= -\Phi' - \frac{3}{r} ,  \\
    \mathcal{K} &= -\frac{1}{r\gamma p} , \\    
    \mathcal{X} &= r(\rho+p)\left[ \omega_{\rm r}^2 e^{2(\Psi-\Phi)} - 8\pi pe^{2\Psi} + \Phi'^2 \right] - 4p' ,  \\
    \mathcal{Y} &= -\Phi' - 4\pi(\rho+ p)re^{2\psi} ,
\end{align}
where $\gamma= (1+\rho/p)v_s^2$ is the adiabatic index. The initial conditions at the center of the DES are $\Delta p\vert_{r=0}= -3\zeta\gamma p\vert_{r=0}$ and $\zeta(0) =1$. Similarly to the nonradial eigenfrequency problem, given a set of test values for $\omega_{\rm r}^2$, the correct frequencies will be those that satisfy the boundary condition $\Delta p= 0$ at $r=R$. Solving this problem yields a discrete spectrum of eigenvalues, where the fundamental mode is the most important because it determines the radial stability and is the first to manifest any instability in the compact star.

The radial oscillation spectrum of our stellar models incorporating DE is shown in Fig.~\ref{FigRadFreq}. The upper panels display the behavior of the squared frequency of the fundamental mode ($\omega_{\rm fr}^2$) as a function of the central energy density for the range $f\in [0,50]\%$. For both the SLy4 and BSR8 EoSs describing the normal matter component, higher oscillation frequencies are obtained when an increasing amount of DE is present in the stellar fluid compared to the purely hadronic case. Furthermore, for larger values of $f$, the critical central energy density at which the transition from stability to instability occurs is shifted to higher values. In other words, the presence of DE inside NSs enhances their radial stability. Our results further indicate that the critical energy density at which $\omega_{\rm fr}^2 = 0$ coincides precisely with the maximum-mass configuration, as in the traditional scenario of pure ordinary matter. The lower panels show the time required for each star to undergo a radial pulsation, namely the period of the fundamental oscillation mode $\tau_{\rm fr} = 2\pi/\omega_{\rm fr}$, which is expressed in milliseconds over the considered range of central densities for our mixed configurations containing DE through the CDF. As the central density increases, the vibration period reaches a minimum and subsequently diverges as the stellar configurations become unstable, irrespective of the percentage fraction~$f$.

%%%%%%%%%%%%%%%%%%%%%%%%%%%%%%%%%%%%%%%%%%%%%%%%%
\section{Summary and future outlook}\label{section4}

Our theoretical model for the EoS of a compact star is the first to be extended beyond ordinary nuclear matter to encompass the effects of DE via a Chaplygin-like fluid. By introducing a parameter $f$, we were able to control the ratio of the DE density to the total energy density of the stellar system. In other words, we constructed a dense stellar fluid that basically contains a mixture of normal matter and DE, finding that increasing the percentage rate of DE leads to a softer total EoS relative to pure nuclear material. This result is important because stiff EoSs with superluminal speed of sound at high densities, such as the APR EoS \cite{APRref}, could be softened in the presence of DE and thus satisfy the causality condition even in the high-density region. Therefore, the causal limit for a nuclear matter EoS can be substantially improved by adding a Chaplygin gas.

We have investigated the impact of DE on the macroscopic properties of NSs when incorporated into their interiors in varying proportions. In particular, our mass--radius predictions indicated that maximum-mass values decrease significantly as a consequence of increasing the percentage rate $f$. This means that our results favor the description of light compact stars and, in that regard, we have shown that the central compact object within the supernova remnant HESS J1731-347 can be consistently described as a DE admixed NS for a wide range of $f$ values. Furthermore, our findings have demonstrated that the presence of DE inside NSs leads to an enhanced gravitational redshift and induces significant modifications in their oscillation properties. Dark energy reduces the tidal deformability, markedly alters both radial and nonradial oscillation spectra, and shifts the onset of radial instability toward higher central densities, thereby increasing stellar stability. These distinctive and potentially observable signatures make oscillation modes, particularly the $f$-mode, powerful probes for discriminating DE admixed NS from purely hadronic configurations. Since the universe is composed of ordinary matter, dark energy and dark matter, this work also paves the way for future research that includes both dark components in addition to normal matter.

\begin{acknowledgments}
This work is part of the project INCT-FNA proc. No. 408419/2024-5. It is also supported by Conselho Nacional de Desenvolvimento Cient\'ifico e Tecnol\'ogico (CNPq) under Grants No. 307255/2023-9 (O.L.), No. 301779/2025-2 (M.D.), No. 401565/2023-8~(Universal - O.L., M.D.), No. 409736/2025-2~(Universal - O.L., M.D.), No. 444797/2024-6 (O.L., M.D.), and Funda\c{c}\~ao de Amparo \`a Pesquisa do Estado de S\~ao Paulo (FAPESP) under Thematic Project No. 2024/17816-8 (O.L., M.D.). J.M.Z.P.  acknowledges support from ``Fundação Carlos Chagas Filho de Amparo à Pesquisa do Estado do Rio de Janeiro'' -- FAPERJ, Process SEI-260003/000308/2024.
\end{acknowledgments}

% The \nocite command causes all entries in a bibliography to be printed out
% whether or not they are actually referenced in the text. This is appropriate
% for the sample file to show the different styles of references, but authors
% most likely will not want to use it.

% \newpage
%

\end{document}